\documentclass[10pt,twocolumn,letterpaper]{article}

\usepackage{cvpr}              

\usepackage{algorithm}
\usepackage{algpseudocode}

\usepackage{graphicx}
\usepackage{amsmath}
\usepackage{amssymb}
\usepackage{booktabs}

\usepackage[pagebackref,breaklinks,colorlinks]{hyperref}

\usepackage[capitalize]{cleveref}
\crefname{section}{Sec.}{Secs.}
\Crefname{section}{Section}{Sections}
\Crefname{table}{Table}{Tables}
\crefname{table}{Tab.}{Tabs.}

\def\cvprPaperID{9176} 
\def\confName{CVPR}
\def\confYear{2022}

\DeclareMathOperator{\SoftMax}{softmax}

\begin{document}

\title{Towards Data-Free Model Stealing in a Hard Label Setting}

\author{\textbf{Sunandini Sanyal} \qquad \textbf{Sravanti Addepalli} \qquad \textbf{R. Venkatesh Babu }\\
Video Analytics Lab, Department of Computational and Data Sciences \\
Indian Institute of Science, Bangalore\\
}

\maketitle

\begin{abstract}

Machine learning models deployed as a service (MLaaS) are susceptible to model stealing attacks, where an adversary attempts to steal the model within a restricted access framework. While existing attacks demonstrate near-perfect clone-model performance using softmax predictions of the classification network, most of the APIs allow access to only the top-1 labels. In this work, we show that it is indeed possible to steal Machine Learning models by accessing only top-1 predictions (Hard Label setting) as well, without access to model gradients (Black-Box setting) or even the training dataset (Data-Free setting) within a low query budget. We propose a novel GAN-based framework\footnote{Project Page: \url{https://sites.google.com/view/dfms-hl}} that trains the student and generator in tandem to steal the model effectively while overcoming the challenge of the hard label setting by utilizing gradients of the clone network as a proxy to the victim's gradients. We propose to overcome the large query costs associated with a typical Data-Free setting by utilizing publicly available (potentially unrelated) datasets as a weak image prior. We additionally show that even in the absence of such data, it is possible to achieve state-of-the-art results within a low query budget using synthetically crafted samples. We are the first to demonstrate the scalability of Model Stealing in a restricted access setting on a 100 class dataset as well.

\end{abstract}
\section{Introduction}
\label{sec:intro}
Deep learning based systems have progressed leaps and bounds over the past few years, enabling their deployment in critical applications such as self-driving cars, surveillance systems and biomedical applications. Furthermore, organizations often provide pretrained machine learning models as a service (MLaaS) where the end user is allowed to query the model and get access to its predictions via APIs for use in various applications.

\begin{figure}
\centering
        \includegraphics[width=\linewidth]{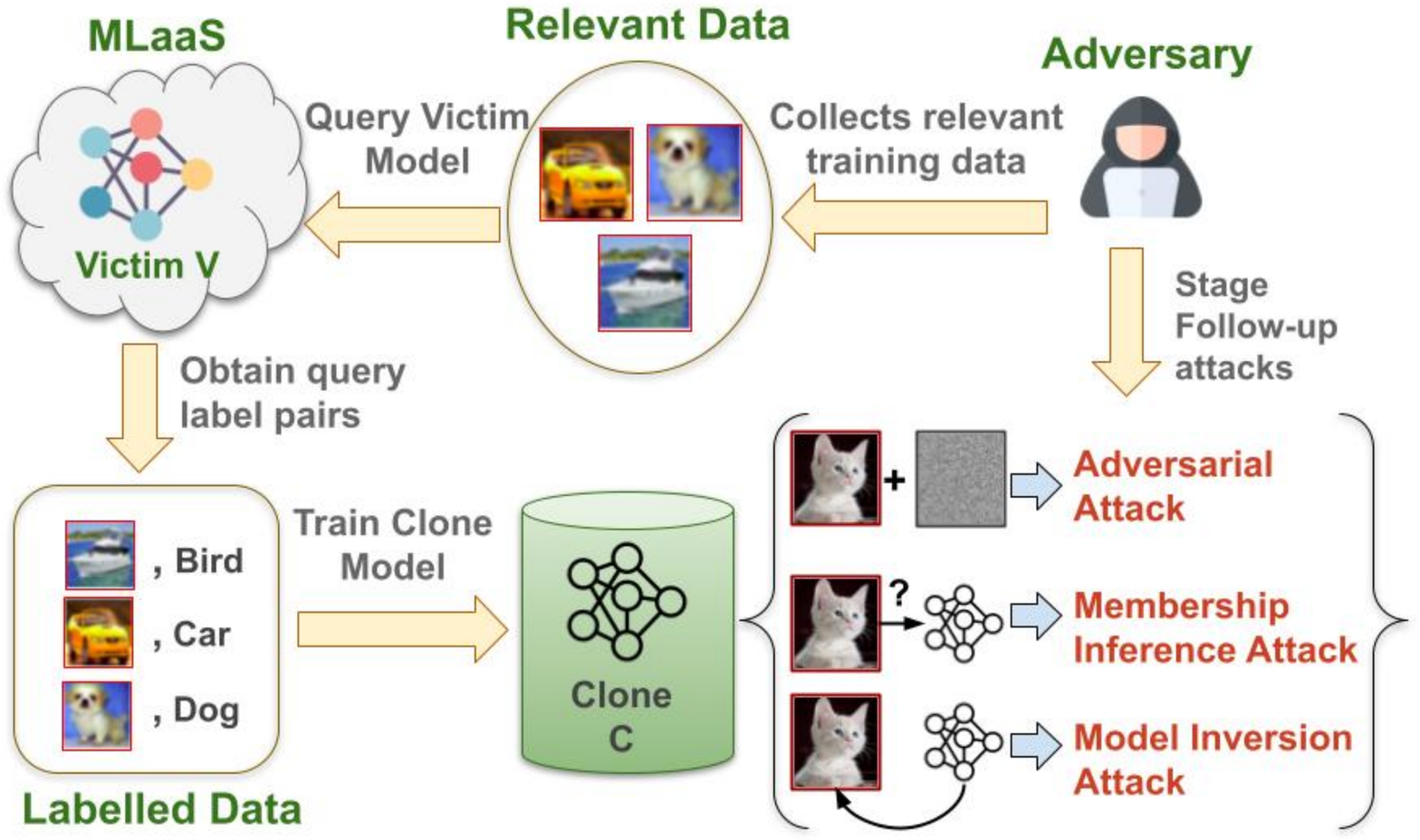}
        \caption{\textbf{Model Stealing Attack and its vulnerabilities}: An adversary queries the victim Model $\mathcal{V}$ with proxy data to obtain its labels. The labelled training data is used to train a clone Model $\mathcal{C}$ which can be further used by the adversary to stage membership inference~\cite{shokri2017membership}, model inversion~\cite{fredrikson2015model} or adversarial attacks~\cite{zhou2018transferable}.} \vspace{-1.0em} 
        \label{fig:motivation}
        
\end{figure}

However, exposing the predictions of the models through queries makes the model susceptible to model stealing attacks, which attempt to clone the victim model even in a black-box setting that restricts access to its gradients. Protecting the privacy of an ML model is of paramount importance as organizations invest significant resources on cutting edge research and also on gathering and labelling large amounts of training data~\cite{halevy2009unreasonable} for achieving competent performance on various tasks. In addition, recent works ~\cite{papernot2017practical, tramer2017space, zhou2020dast, wang2021delving} have shown that an adversary could train a substitute model via model stealing and use it for crafting adversarial examples~\cite{goodfellow2014explaining} in a black-box setting, which poses a serious threat when the model is deployed in security critical applications. A stolen model could also compromise the privacy of users by leaking confidential data through a membership inference attack~\cite{shokri2017membership} or model inversion~\cite{zhang2020secret, zhao2021exploiting}. Fig.-\ref{fig:motivation} showcases some of the possible malicious outcomes of Model Stealing. In order to prevent model stealing attacks, some defenses attempt to perturb the softmax predictions of the model, while preserving its top-1 prediction~\cite{lee2018defending}. In this work we consider the problem of model stealing in a more practical and challenging hard label setting, where only the top-1 prediction of the model is accessible, and is thus effective even in the presence of such defenses. 

In a model stealing attack, an adversary first queries a black-box victim model $\mathcal{V}$ with input data and obtains a prediction for it as shown in Fig.\ref{fig:motivation}. This data along with victim model predictions is used to train a clone model $\mathcal{C}$. In a practical scenario, the attacker would not have access to the training data, and hence we consider the problem of Data-Free Model Stealing (DFMS) in this work. In such a data-free scenario, the attacker could use publicly available related datasets \cite{papernot2017practical,orekondy2019knockoff}, or synthetically generated samples \cite{truong2021data} to query the model. While the use of publically available datasets assumes access to related data, the data-free generative approach could suffer from a large query budget, as the synthetic data can be far from the true training data distribution. In this work, we overcome both challenges by utilizing the available data that may be unrelated to the original training dataset, as a weak image prior. This enables the generation of representative samples under a low query budget, which is a crucial requirement in model stealing attacks, since MLaaS APIs work on a pay-per-query basis. 

While existing algorithms for Data-Free Knowledge Distillation~\cite{addepalli2020degan, nayak2019zero, lopes2017data, yin2020dreaming, fang2019data} and Model Extraction~\cite{kariyappa2021maze,truong2021data} achieve near perfect clone-model accuracy, there are additional challenges in a Model Stealing framework due to the lack of access to gradients and a hard-label setting. Therefore, we consider a practical setup of data-free hard-label model stealing and overcome the challenges by utilizing the clone model’s gradients as a proxy to the gradients of the victim model. As the clone model starts training, it acts as a useful proxy for the victim model, and helps the generator learn to generate rich informative samples, which boosts the clone accuracy further. We explicitly enforce the generation of a class-balanced dataset from the generator that is also more aligned with the distribution of the training dataset. 
 
Additionally, we also utilize an adversarial loss in a GAN framework \cite{goodfellow2014generative}, by using publicly available potentially unrelated data, which we refer to as proxy data \cite{addepalli2020degan}. While this could be completely unrelated to the original training dataset, it still helps in enforcing a weak image prior in the generated data. This in turn reduces the number of victim model queries needed to perform Model Stealing. In fact, we show that it is possible to even use synthetic samples, such as multiple overlapping shapes with a planar background, to steal a model in a completely data-free setting. 

Our method achieves a significant improvement over  ZSDB3KD \cite{wang2021zero}, a zero-shot data-free method in a similar hard label setting using only synthetic samples. In the upcoming sections, we describe our approach in detail and show results on various datasets. 

\noindent Our \textbf{key contributions} are listed below: 
\begin{itemize}
\itemsep0em
\item We propose  DFMS-HL, a Data-Free Model Stealing (DFMS) attack in a Hard-Label (HL) setting, to train a clone model with the help of unrelated proxy data or manually crafted synthetic data. 
We show that DFMS-HL outperforms the existing baseline ZSDB3KD \cite{wang2021zero} and results in a significant reduction of around $500\times$ in the number of queries to the victim model.
\item We demonstrate state-of-the-art results on the CIFAR-10 dataset 
using unrelated proxy samples, such as a given subset (containing 40 or 10 non-overlapping classes) from CIFAR-100, 
or synthetic data.
\item We are the first to show noteworthy results of data-free model stealing on a dataset with a larger number of classes such as CIFAR-100. This demonstrates that our approach is both effective and scalable.  
\item We compare our method with the state-of-the-art model stealing attacks MAZE~\cite{kariyappa2020protecting} and DFME~\cite{truong2021data}, which additionally utilize softmax predictions of the victim model. Although we consider a more restrictive setting, we achieve a comparable accuracy using the DFMS-HL approach, and a significant boost of around 3\% using a Soft-Label (SL) variant of the proposed method (DFMS-SL).
\end{itemize}

\section{Related Work}

In this section, we discuss existing Knowledge Distillation and Model Stealing works with varied levels of access to the victim model as shown in Table-\ref{table:taxonomy}. 
\subsection{Knowledge distillation} 
Knowledge distillation~\cite{hinton2015distilling} aims to transfer the knowledge of a large pretrained teacher model to a smaller student model without a significant impact on accuracy. This is primarily used to compress models for deployment, in order to reduce the memory requirements and inference time~\cite{gou2021knowledge, adriana2015fitnets, yang2020model, aguinaldo2019compressing}. In practical scenarios, training data is kept confidential due to privacy concerns. Hence, there has been a lot of focus on developing data-free approaches for knowledge-distillation. ZSKD~\cite{nayak2019zero}, DAFL~\cite{chen2019data}, DFKD~\cite{lopes2017data} are popular knowledge distillation methods in a data-free setting. A data-free KD method DeGAN~\cite{addepalli2020degan} demonstrated that it is possible to use publicly available unrelated data (proxy dataset) to distill the knowledge of a teacher model to a smaller student model. However, all these methods require access to the teacher model’s gradients. Following this, Black-Box Ripper~\cite{barbalau2020black} was proposed to implement model stealing by querying a black-box teacher model with unrelated proxy data. A recent work ZSDB3KD~\cite{wang2021zero} proposed knowledge distillation for a black box model with only hard-label outputs. However, this approach is highly computationally intensive due to the requirement of a very large number of queries (4000 million) to the teacher model. Our work considers the same setup of having access to only the top-1 labels, with a significantly lower query budget of 8 million.
\begin{table}
\caption{Taxonomy of prior works on Knowledge Distillation (KD) and model stealing attacks. Our approach DFMS-HL is a data-free model stealing attack on a black-box victim model with access to only hard labels.}
\setlength\tabcolsep{3pt}
\centering
\resizebox{0.95\linewidth}{!}{
\label{table:taxonomy}
\begin{tabular}{cccc}
\toprule
\textbf{Approach} & \textbf{\begin{tabular}[c]{@{}c@{}}White-Box \\ Soft Label\end{tabular}} & \textbf{\begin{tabular}[c]{@{}c@{}}Black-Box \\ Soft Label\end{tabular}} & \textbf{\begin{tabular}[c]{@{}c@{}}Black-Box \\ Hard Label\end{tabular}} \\
\midrule
Data free         & \begin{tabular}[c]{@{}c@{}}ZSKD~\cite{nayak2019zero}\\ DeGAN~\cite{addepalli2020degan}\end{tabular}                     & \begin{tabular}[c]{@{}c@{}}MAZE~\cite{kariyappa2021maze}\\ DFME~\cite{truong2021data}\end{tabular}                      & \begin{tabular}[c]{@{}c@{}}ZSDB3KD~\cite{wang2021zero}\\  DFMS-HL (Ours)\end{tabular}               \\
\midrule
Data              & \begin{tabular}[c]{@{}c@{}}KD with\\ Data~\cite{hinton2015distilling}\end{tabular}                & \begin{tabular}[c]{@{}c@{}}KnockoffNets~\cite{orekondy2019knockoff}\\ JBDA~\cite{papernot2017practical}\end{tabular}              & -    \\ 
\bottomrule
\end{tabular}}
\vspace{-0.3cm}
\end{table}
\subsection{Model Stealing}

Tramer \etal \cite{tramer2016stealing} demonstrated that an attacker could use queries to steal a machine learning model with near perfect fidelity. Following this, model stealing has been implemented in various domains ~\cite{krishna2019thieves, jagielski2020high, pal2019framework, correia2018copycat, milli2019model}. A partial data approach JBDA~\cite{papernot2017practical} assumed access to a small set of samples from the data distribution. On the other hand, surrogate data approaches such as KnockOffNets ~\cite{orekondy2019knockoff} and Black-Box dissector\cite{wang2021black}  consider that attackers could use images from a different data source to steal a model. These methods fail to perform well without a suitable surrogate dataset. 
This motivated the development of data-free approaches which work well without using surrogate data or seed samples from the training data. Recent data-free approaches such as MAZE~\cite{kariyappa2021maze} and DFME~\cite{truong2021data} attempt to extract models using GAN generated synthetic data. In these approaches, the generator is trained to produce images that maximize the dissimilarity score between the clone and victim models. The victim model's gradients are required to measure this dissimilarity score, and are estimated using zeroth-order gradient approximation. These approaches are computationally expensive as they require a lot of queries ($\sim$20 million) to the victim model for synthesizing data samples in a black-box setting.  Moreover, these  methods assume that the  softmax vector from the teacher model is accessible. Contrary to this, we consider a practical setting that allows access to only hard labels from the victim model.

\subsection{Defenses against model stealing}
Lee \etal \cite{lee2018defending} propose to defend against model stealing attacks by perturbing the model predictions while preserving its top-1 label, to maintain similar classification accuracy. 
On similar lines, Prediction Poisoning~\cite{orekondy2019prediction} perturbs model predictions by poisoning the output distribution at the cost of model accuracy. However, such defenses fail in a scenario where an attacker has access to only hard labels from the model. A more sophisticated approach EDM~\cite{kariyappa2020protecting} introduces randomness into the predictions by using an ensemble of diverse models to produce dissimilar outputs for Out-of-Distribution (OOD) samples, that are likely to be used for querying the victim model in a model stealing attack. Similarly, Adaptive Misinformation~\cite{kariyappa2020defending} perturbs the predictions for OOD inputs only. However, these approaches have been shown to cause utility degradation \cite{orekondy2019prediction}, or can be made ineffective using an adaptive query synthesis strategy \cite{chandrasekaran2020exploring}. Further, Chandrasekaran \etal~\cite{chandrasekaran2020exploring, chandrasekaran2021sok} provide theoretical insights to demonstrate that ``model extraction is inevitable", even in a realistic setting with only hard labels, and even when models use randomised defenses. Hence, a model with a reasonably good accuracy would always leak information that could lead to model extraction. In this work we demonstrate that it is indeed possible to perform model stealing in a severely restricted setting as well, and further achieve competent clone accuracy. This paves way to the development of better defenses for preserving model privacy in future. 

\section{Proposed Approach}

For model stealing, the goal of an adversary is to learn the parameters of the clone model $\mathcal{C}$ so as to match the predictions of the victim model $\mathcal{V}$. Towards this end, we propose a data-free model stealing approach \textbf{DFMS-HL} that requires only hard-label access. In the following sections, we describe the proposed model stealing attack algorithm.

\begin{figure*}[ht]
\centering
        \includegraphics[width=0.80\linewidth]{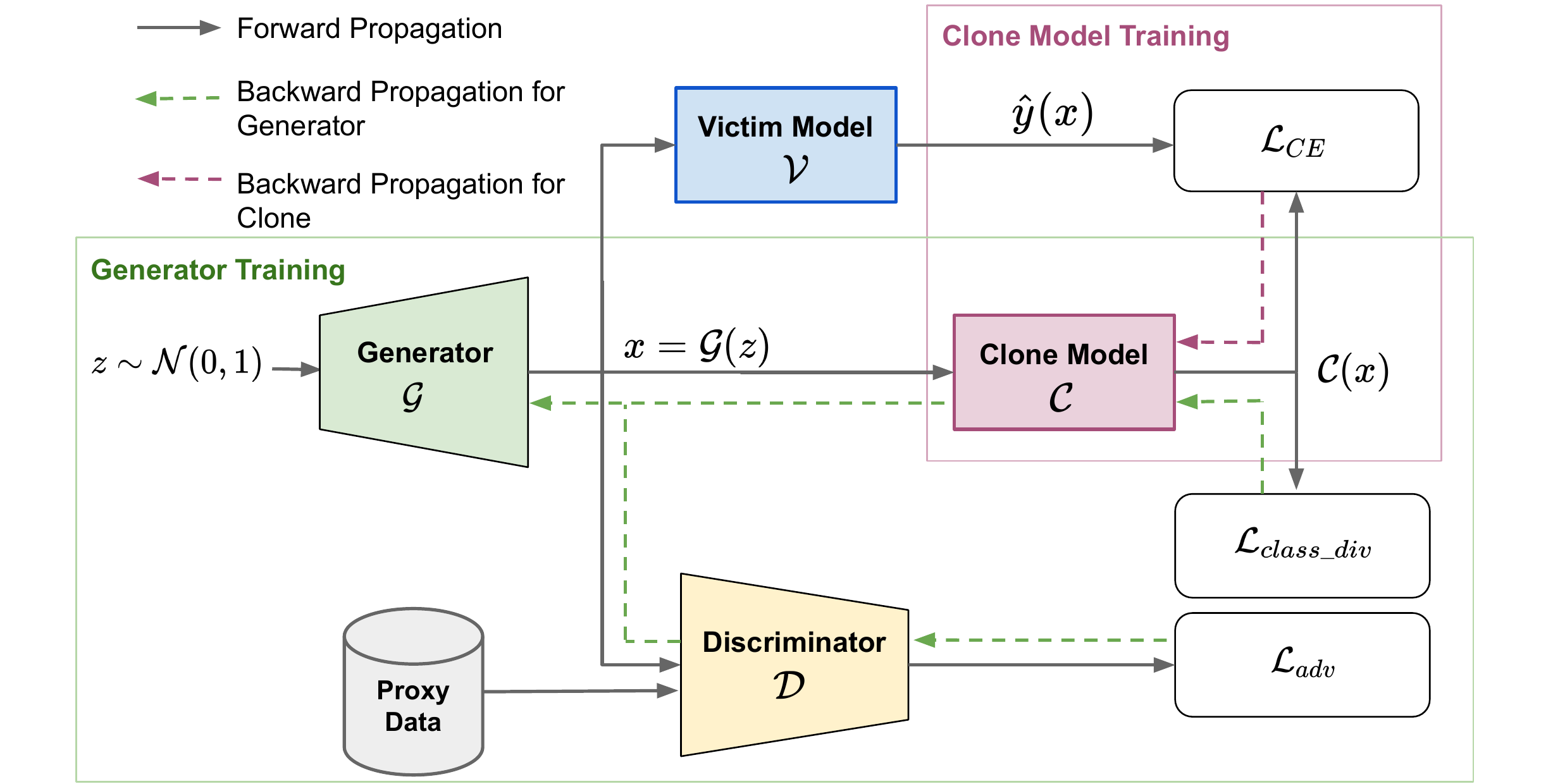}
        \caption{\textbf{Architecture of DFMS-HL}: Generator $\mathcal{G}$ generates data $x$ with a proxy image prior. The clone model $\mathcal{C}$ is trained using the predictions from the victim model $\mathcal{V}$ with a cross-entropy loss objective $\displaystyle \mathcal{L}_{CE}$. The discriminator $\mathcal{D}$ learns to discriminate between proxy data and the samples generated from $\mathcal{G}$. The generator $\mathcal{G}$ is trained using the adversarial loss $\mathcal{L}_{adv}$ along with the class-diversity loss $\mathcal{L}_{class\_div}$. The generator and clone model are trained alternately in every iteration of the algorithm.}  
        \label{fig:architecture}
\end{figure*}

\subsection{Overview}

We use a GAN based architecture to train the clone model. We first train a DCGAN\cite{radford2015unsupervised} by imposing an image prior using synthetic data or unrelated proxy data, and use this as an initialization for the generator $\mathcal{G}$. Further, the clone model and generator are trained alternately. The data flow of the proposed model stealing attack is shown in Fig.~\ref{fig:architecture}, wherein the generator $\mathcal{G}$ generates data $x=\mathcal{G}(z)$ from a random vector $z$. The victim model takes input $x$ and generates input-label pairs $(x,\hat{y}(x))$. Since, the victim model is black-box, we do not backpropagate the gradients through it. We use the input-label pairs to train the clone model. Further, the generated data $x$ is used to train the generator using the adversarial loss \cite{goodfellow2014generative} and a diversity loss \cite{addepalli2020degan}. The discriminator learns to differentiate between fake and real proxy data using the adversarial loss. In the subsequent sections, we describe the loss functions for training the generator and clone model in further detail.

\subsection{Clone model Training}
The clone model $\mathcal{C}$ is trained using the data samples generated from the generator $\mathcal{G}$. In every iteration, we sample an $m$-dimensional random vector $z$, whose elements are sampled from $m$ \textit{i.i.d.} Standard Normal distributions. This vector is forward propagated through $\mathcal{G}$ to generate images $x$. These images are then passed to the victim model to obtain its hard-labels. The clone model is trained with the cross-entropy loss objective using the victim predictions as ground truth, as shown below:
\vspace{-0.5em}
\begin{equation}
\mathcal{L}_{C} = \underset{z \sim \mathcal{N}(0,I)}{\mathbb{E}} \left[ \mathcal{L}_{CE}(\mathcal{C}(x),\hat{y}(x)) \right], 
\ 
x=\mathcal{G}(z)
\end{equation}
where
$\displaystyle \ \hat{y}(x) = \underset{i}{\mathrm{argmax}}\ \mathcal{V}_i(x) $ is the class label corresponding to the maximum probability class, $I$ is an $m$ dimensional identity matrix, and $\mathcal{C}(x)$ is the pre-softmax output from the clone model.

\begin{algorithm}[t]
\caption{DFMS-HL : Algorithm for Model Stealing}\label{alg:MoSAlgo}
\begin{algorithmic}
\Require $N_Q, \mathcal{G}, \mathcal{D}, n_G, n_C $
\State // Initialize a Generator $\mathcal{G}$ with DCGAN parameters
\State // Train the clone model $\mathcal{C}$ with DCGAN and proxy images using $n_C$ queries for initialization.
\While{$n_G \neq 0$}
    \State $x = \mathcal{G}(z),  z\sim \mathcal{N}(0,I)$
    \State $\mathcal{L}_G \gets
    \mathcal{L}_{adv, fake} + \lambda_{div} \mathcal{L}_{class\_div}$ 
    \State $\mathcal{L}_D \gets  \mathcal{L}_{adv,real} + \mathcal{L}_{adv,fake}$
    \State $\mathcal{\theta_G} \gets \mathcal{\theta_G} - \mathcal{\epsilon_G} \nabla_{\theta_G}\mathcal{L}_G$
    \State $\mathcal{\theta_D} \gets \mathcal{\theta_D} - \mathcal{\epsilon_D} \nabla_{\theta_D}\mathcal{L}_D$
    \State $n_G \gets n_G-1$
\EndWhile
\State // Train clone model $\mathcal{C}$ 
\While{$n_C \neq 0$}
    \State $x = \mathcal{G}(z),  z\sim \mathcal{N}(0,I)$
    \State $\mathcal{L}_{C} \gets  \mathcal{L}_{CE} (\mathcal{C}(x), \hat{y}(x)) $
    \State $\mathcal{\theta_C} \gets  \mathcal{\theta_C} - \mathcal{\epsilon_C} \nabla_{\theta_C}\mathcal{L}_C$
    \State $n_C \gets n_C-1$
\EndWhile
\State // Start alternate training between $\mathcal{G}$ and $\mathcal{C}$
\While{$N_Q \neq 0$}

    // Train $\mathcal{G}$ and $\mathcal{D}$ with $\mathcal{C}$ as fixed
    \State $x = \mathcal{G}(z),  z\sim \mathcal{N}(0,I)$
    \State $\mathcal{L}_G \gets
    \mathcal{L}_{adv, fake} + \lambda_{div} \mathcal{L}_{class\_div}$ 
    \State $\mathcal{L}_D \gets \mathcal{L}_{adv,real} + \mathcal{L}_{adv,fake}$
    \State $\mathcal{\theta_G} \gets  \mathcal{\theta_G} - \mathcal{\epsilon_G} \nabla_{\theta_G}\mathcal{L}_G$
    \State $\mathcal{\theta_D} \gets  \mathcal{\theta_D} - \mathcal{\epsilon_D} \nabla_{\theta_D}\mathcal{L}_D$

// Train $\mathcal{C}$ with $\mathcal{G}$ and $\mathcal{D}$ as fixed
    \State $x = \mathcal{G}(z),  z\sim \mathcal{N}(0,I)$
    \State $\mathcal{L}_{C} \gets  \mathcal{L}_{CE} (\mathcal{C}(x), \hat{y}(x)) $
    \State $\mathcal{\theta_C} \gets \mathcal{\theta_C} - \mathcal{\epsilon_C} \nabla_{\theta_C}\mathcal{L}_C$
\EndWhile

\end{algorithmic}
\end{algorithm}

\subsection{Generator Training}
For imposing an image prior, we initially train a DCGAN generator using proxy data or synthetic images. However, we find that this is not sufficient as the generator could potentially suffer from mode collapse and lack of diversity. Moreover, lack of class diversity can severely impact the learning of tail classes in a hard-label setting.
Hence, it crucial for the generator to generate a class-balanced set of images for learning the information across all classes. Therefore, we use a class-diversity loss formulation \cite{addepalli2020degan} to generate diverse samples from the generator $\mathcal{G}$ while remaining close to the manifold of the proxy/synthetic images. 

The generator loss has two components. The first component is the adversarial loss \cite{goodfellow2014generative} which causes the generator to generate data close to the proxy data distribution. The second component is a class balancing loss \cite{addepalli2020degan}, to enforce a diversity constraint. 
The two loss formulations for the generator are described in more detail below. 

\textbf{Adversarial Loss}~\cite{goodfellow2014generative}: The adversarial loss ensures that the distribution of images is close to the images in the  proxy or synthetic dataset.
\vspace{-1.0em}
    \begin{equation}
        \displaystyle \mathcal{L}_{adv,real} = \underset{x\sim p_{data}(x)}{\mathbb{E}} \left[  log \mathcal{D}(x) \right]
    \end{equation}
    \begin{equation}
        \displaystyle \mathcal{L}_{adv,fake} = \underset{z\sim \mathcal{N}(0,I)}{\mathbb{E}} \left[  log (1 - \mathcal{D}(\mathcal{G}(z)) \right]
    \end{equation}
    
    The discriminator $\mathcal{D}$ and generator $\mathcal{G}$ play a min-max game~\cite{goodfellow2014generative} as follows:
    \begin{equation}
        \displaystyle \underset{\mathcal{G}}{min}\  \underset{\mathcal{D}}{max} \  \mathcal{L}_{adv,real} + \mathcal{L}_{adv,fake}
    \end{equation}
    
    \textbf{Class Diversity Loss \cite{addepalli2020degan}}: The class diversity loss encourages the generation of diverse images across all classes. In a batch of $N$ samples, we consider the expected confidence value over the batch as $\alpha_j$ for every class $j$, and obtain the entropy over all $K$ classes. The negative entropy, denoted as $\mathcal{L}_{class\_div}$ is computed as shown below:
    \vspace{-1.0em}
    \begin{equation}
        \displaystyle \mathcal{L}_{class\_div} =    \sum_{j=0}^{K} \alpha_j \log \alpha_j
    \end{equation}
    \vspace{-0.5em}
    \begin{equation}
        \displaystyle \alpha_j =  \frac{1}{N} \sum_{i=1}^{N} \SoftMax(\mathcal{C}(x_i))_j
    \end{equation}

    \textbf{Using Clone Model as a Proxy for Victim:} Since, the victim model is black-box, backpropagation through $\mathcal{V}$ is not permitted. Hence, for imposing diversity we use the clone model parameters to compute the loss. Over the training process, the clone learns to imitate the gradients of the victim, making it a suitable proxy for enforcing diversity in the generated images. 

The equations given below describe the overall generator and discriminator losses. 
\begin{equation}
   \mathcal{L}_G  =  \mathcal{L}_{adv, fake} + \lambda_{div}  \mathcal{L}_{class\_div}  
\end{equation}
\begin{equation}
    \mathcal{L}_D = \mathcal{L}_{adv,real} + \mathcal{L}_{adv,fake}
\end{equation}

\subsection{Algorithm}
The overall training algorithm is outlined in Algorithm-\ref{alg:MoSAlgo}. We first train a DCGAN to initialize the generator model with an image prior. Following this, we train the clone model using a mix of images from the DCGAN and the proxy dataset to obtain a good initialization for the clone model. Using this clone model, we further fine-tune the generator for $n_G$ epochs using the two proposed losses; adversarial loss and class-diversity loss. We then train a clone model from scratch for $n_C$ epochs using the images from the diverse generator $\mathcal{G}$. Following this, we start the alternate training process for the generator and clone model. We train the generator for one iteration by freezing weights of the clone model and subsequently train the clone model for one iteration using labels from the victim model. This procedure is repeated until the query budget $N_Q$ is exhausted.

\subsection{Computing the Query Cost}

In this section, we compute the total number of queries to the victim model. The number of samples in the proxy data is denoted as $N_P$. Initially, we require $n_C$ queries to obtain a clone model to initialize the generator and an additional $n_C$ queries to initialize the Classifier $\mathcal{C}$. For our experiments, we set $n_C$ as 50,000. The alternate training of the clone and generator continues for $E$ epochs and in each epoch, the victim model is queried $N_P$ times. So the total query cost is computed as follows,
\vspace{-0.3cm}
\begin{equation}
    N_Q = E \cdot N_P
\end{equation}
\vspace{-0.4cm}
\begin{equation}
    \displaystyle \mathrm{Total\ Queries} = 2 \cdot n_C + N_Q
\end{equation}
We set the query limit $N_Q$ to 8 million for our proxy and synthetic data experiments on CIFAR-10.

\subsection{Insights on Query Budget}
Chandrasekaran \etal\cite{chandrasekaran2020exploring} formulated the model extraction task as a query synthesis active learning problem where an adversary learns a hypothesis function with a query complexity $q_A(\epsilon, \delta)$. The authors observe that it is possible for an adversary to implement an $\epsilon$-extraction attack with query complexity $q_A(\epsilon, \delta)$ and confidence $1-\delta$ (described in Section 2 of the Supplementary). The authors\cite{chandrasekaran2020exploring} further prove that model stealing is inevitable and there exists a query bound within which a model could be stolen. We empirically find the query budget needed for the proposed approach in the Query ablation (Section \ref{sec:abla_study}).

\begin{table*}
\caption{\textbf{ Comparison of DFMS-HL with state-of-the-art KD methods (Top) and ZSDB3KD (Bottom) on CIFAR-10}: Clone model accuracy (\%) reported using Proxy data as unrelated (40 or 10) CIFAR-100 classes and synthetic data. Victim and clone model architectures used are Alexnet and AlexNet-half respectively.}
\setlength\tabcolsep{3pt}
\centering
\resizebox{0.99\linewidth}{!}{
\label{table:cifar10_main}
\begin{tabular}{lccccccc}
\toprule
Method  & Hard Label & Black-Box & Data-Free & Victim Accuracy & Synthetic/ Data-Free      & CIFAR-100 (40C) & CIFAR-100 (10C) \\
\midrule
\multicolumn{8}{c}{\small{\textbf{Victim Accuracy = 82.5\%}}} \\
\midrule
ZSKD~\cite{nayak2019zero}             & $\times$                  & $\times$                 & $\checkmark$                 & 82.50                & \textbf{69.50}          & -           & -           \\
DeGAN~\cite{addepalli2020degan}            & $\times$                  & $\times$                 & $\checkmark$                 & 82.50                & -              & 76.30           & 72.60           \\
KnockoffNets~\cite{orekondy2019knockoff}     & $\times$                  & $\checkmark$                 & $\checkmark$                 & 82.50                & -              & 65.70           & 46.60           \\
Black-Box Ripper~\cite{barbalau2020black} & $\times$                  & $\checkmark$                 & $\checkmark$                 & 82.50                & -              & \textbf{76.50}           & \textbf{77.90}           \\
DFMS-HL (Ours)   & $\checkmark$                   & $\checkmark$                 & $\checkmark$                 & 82.52                & 65.70          & 76.02               & 71.36           \\
\midrule
\multicolumn{8}{c}{\small{\textbf{Victim Accuracy $\sim$ 80\%}}} \\
\midrule
ZSDB3KD~\cite{wang2021zero}          & $\checkmark$                 & $\checkmark$                & $\checkmark$                & 79.30                & 59.46          & 59.46           & 59.46           \\
DFMS-HL (Ours)   & $\checkmark$                 & $\checkmark$                & $\checkmark$                & 80.18                & \textbf{67.03} & \textbf{74.27}  & \textbf{70.57} \\
\bottomrule
\vspace{0.2em}
\end{tabular}}
\vspace{-0.3cm}
\end{table*}
\begin{table*}
\caption{\textbf{Comparison of DFMS-HL with data-free model stealing methods MAZE, DFME (Top) and ZSDB3KD (Bottom) on CIFAR-10}: Clone Accuracy (\%) is reported using proxy data from unrelated classes (40 or 10) of CIFAR-100 and synthetic data, with victim models as ResNet34 and ResNet-18 for the top and bottom sections respectively. ResNet18 architecture is used for the clone model.}
\setlength\tabcolsep{3pt}
\resizebox{0.99\linewidth}{!}{
\label{table:resnet_main}
\begin{tabular}{lccccccc}
\toprule
Method & Hard Label & Black-Box & Data-Free & Victim Accuracy & Synthetic/ Data-Free & CIFAR-100 (40C) & CIFAR-100 (10C) \\
\midrule
\multicolumn{8}{c}{\small{\textbf{Victim Accuracy $\sim$ 95.5\%, Victim Model: ResNet-34}}} \\
\midrule
MAZE~\cite{kariyappa2020protecting}            & $\times$                  & $\checkmark$                & $\checkmark$                & 95.50                 & 45.60     & -           & -           \\
DFME~\cite{truong2021data}            & $\times$                  & $\checkmark$                & $\checkmark$                & 95.50                 & 88.10     & -           & -           \\
DFMS-HL (Ours)  & $\checkmark$                 & $\checkmark$                & $\checkmark$                & 95.59                 & 84.51     & \textbf{92.06}           & \textbf{85.53}           \\
DFMS-SL (Ours) & $\times$                 & $\checkmark$                & $\checkmark$ &
  95.59 & \textbf{91.24}     & \textbf{93.96}                                                      & \textbf{90.88} \\
\midrule
\multicolumn{8}{c}{\small{\textbf{Victim Accuracy $\sim$ 93.7\%, Victim Model: ResNet-18}}} \\
\midrule
ZSDB3KD~\cite{wang2021zero}         & $\checkmark$                 & $\checkmark$                & $\checkmark$                & 93.65                & 50.18     & -           & -           \\
DFMS-HL (Ours)  & $\checkmark$                 & $\checkmark$                & $\checkmark$                & 93.83                & \textbf{85.92}     & \textbf{90.51}           & \textbf{83.37}          \\
\bottomrule
\end{tabular}}
\vspace{-0.3cm}
\end{table*}

\section{Experiments}

We perform experiments to evaluate the effectiveness of the proposed algorithm DFMS-HL in a hard-label data-free setting. We primarily compare our approach to the existing method ZSDB3KD~\cite{wang2021zero}, which is a zero-shot hard-label Knowledge distillation method. We present evaluations of DFMS-HL by using various proxy datasets as well as with synthetically crafted data. Our attack not only outperforms ZSDB3KD by a large margin, but also achieves clone-model accuracy comparable to the soft-label methods by using only hard-labels from the victim model. Additionally, we perform ablations to highlight the number of queries required to successfully steal a model, and also to understand the impact of the class-diversity loss. Our analysis reveals that the proposed attack is computationally more efficient when compared to existing approaches since it requires significantly lesser queries.

\subsection{Experimental Setup}
We evaluate DFMS-HL on two datasets, CIFAR-10 and CIFAR-100. For evaluation, we first train a victim model with the same teacher accuracy as ZSDB3KD ~\cite{wang2021zero} for a fair comparison. The victim models are trained until the accuracy reaches the expected value. We evaluate our approach using the following two (Victim, Clone) pairs: (ResNet18, ResNet-18) and (AlexNet, AlexNet-half). 

For the clone model training, we use an SGD optimizer with momentum of 0.9, maximum learning rate of 0.1 and a weight decay of $5\times10^{-4}$. We use a cosine annealed scheduler to decay the learning rate across epochs. For initialization, the clone model is trained for 200 epochs. For the main approach, the clone model is further trained with the images generated from the generator within the query budget or until the accuracy saturates.  

For the generator, we use a DCGAN \cite{radford2015unsupervised} with upto five transpose convolution layers followed by batch-normalization and ReLU units. We use Tanh activation units after the last convolution layer to generate images in the normalised range of $[-1,1]$. The discriminator contains a stack of five convolution layers followed by batch normalization and Leaky ReLU units. The last layer of the discriminator uses Sigmoid activation. The GAN is trained with an Adam optimizer~\cite{kingma2014adam} and a learning rate of $2\times10^{-4}$ with $(\beta_1, \beta_2)$ set to $(0.5, 0.999)$.

\subsection{Results}
\textbf{Comparison with Knowledge distillation (KD) methods:} We compare the proposed approach with existing KD methods on CIFAR-10 in Table \ref{table:cifar10_main}. DeGAN~\cite{addepalli2020degan} and ZSKD~\cite{nayak2019zero} are data-free KD methods with white-box teacher access, while KnockoffNets~\cite{orekondy2019knockoff} and Black-Box Ripper~\cite{barbalau2020black} are data-free KD methods in a black-box setting. Similar to the experimental setting of prior works \cite{addepalli2020degan,barbalau2020black}, we use images from 40 unrelated classes of CIFAR-100 as the proxy dataset for CIFAR-10 model stealing. We also show results using images from 10 classes randomly sampled from these 40 unrelated classes. We achieve results comparable to the data-free KD methods despite having more restrictions on access to the victim model. 

\begin{figure}
\centering
     \includegraphics[width=0.78\linewidth]{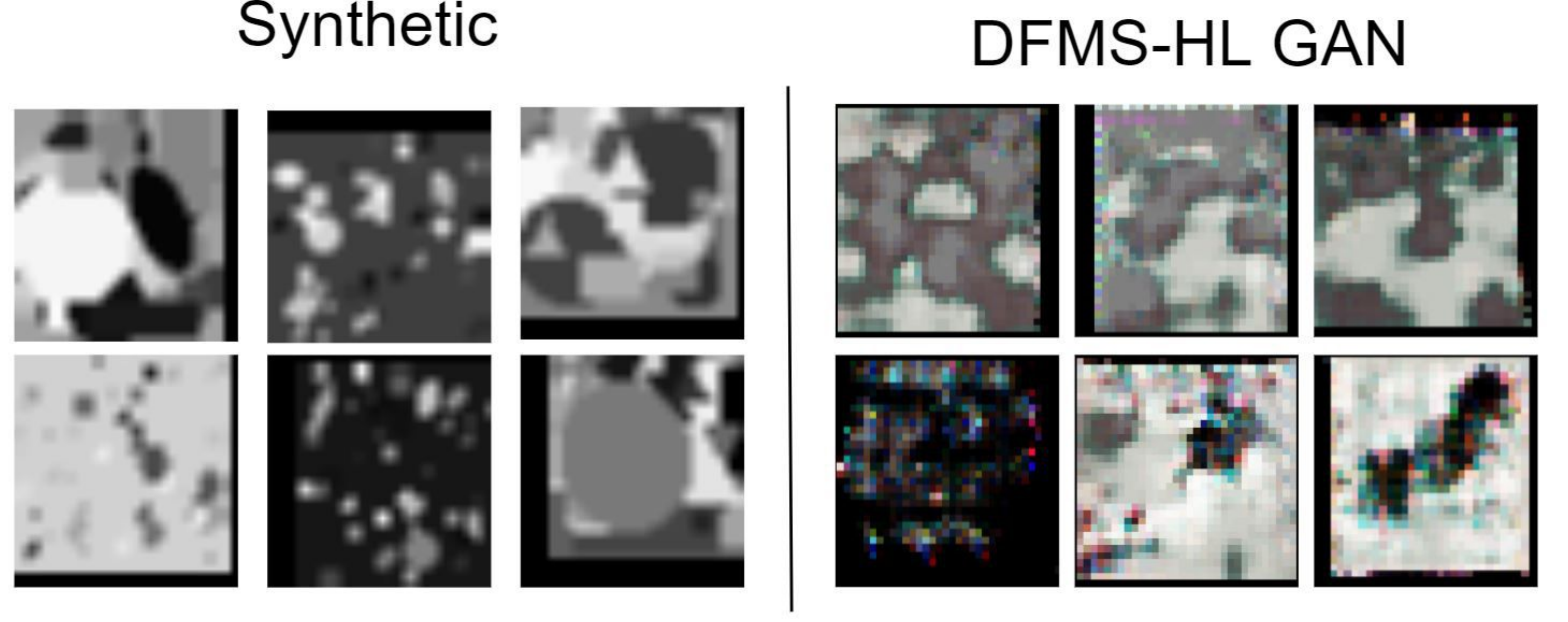}
        \caption{Samples of grey-scale synthetic images shown on the left, along with images generated from the DFMS-HL generator shown on the right.}
        \vspace{-1em}
       \label{fig:gan_imgs}
\end{figure}

We also show results by using synthetically crafted data for imposing image priors using the discriminator. For this, we generate a synthetic dataset of 50k samples by including random shapes (triangle, rectangle, ellipse or circles) of randomly sampled sizes at random locations on a plain background of random color (details in Section 1.1 of the Supplementary). We also generate textured images by increasing the maximum number of shapes to 100 and reducing the maximum region occupied by the shapes in the image. These images are converted to grey-scale as shown in Fig.\ref{fig:gan_imgs}, and further used as proxy data to train the generator. For comparing our results with ZSDB3KD, we train a victim model with a comparable accuracy of 80.18\%. From Table \ref{table:cifar10_main}, it can be observed that our approach not only outperforms ZSDB3KD by a large margin, but also achieves a comparable accuracy with respect to DeGAN and Black-Box Ripper using 40 unrelated classes from CIFAR-100 as the proxy data. We use a significantly lower query budget of 8M when compared to ZSDB3KD which requires 4000M queries. We report the clone model accuracy with other proxy datasets in Table \ref{table:other_proxy_data}. When synthetic data is used, we report our numbers under the ``Data-Free" column across all tables since we do not use any additional data in this case. We obtain significant gains when compared to ZSDB3KD across different proxy datasets.
\begin{table}
\caption{\textbf{Performance of DFMS-HL on CIFAR-100}: Clone Accuracy (\%) achieved on CIFAR-100 with different proxy data. Victim and Clone architectures are ResNet18.}
\setlength\tabcolsep{3pt}
\centering
\resizebox{1\linewidth}{!}{
\label{table:cifar100_main}
\begin{tabular}{lc|cc}
\toprule
Method     & Proxy Data & Victim Accuracy & Clone Accuracy \\
\midrule
DeGAN~\cite{addepalli2020degan}               & CIFAR-10            & 78.52                & 75.62              \\
DFMS-HL (Ours)  & CIFAR-10            & 78.52                & 72.83                  \\
\midrule
DFMS-HL (Ours) & Synthetic           & 78.52                & 43.56             \\
\bottomrule
\end{tabular}}
\vspace{-0.3cm}
\end{table}
\begin{table}
\caption{\textbf{ Clone model accuracy (\%) using DFMS-HL with different proxy datasets}. ResNet-18 architecture is used for both victim and clone models.}
\setlength\tabcolsep{3pt}
\centering
\resizebox{1.0\linewidth}{!}{
\label{table:other_proxy_data}
\begin{tabular}{c|ccccc|cc}
\toprule
Victim training Data:  & \multicolumn{5}{c|}{CIFAR-10}                                                                                  & Fashion MNIST\\
\midrule
Proxy Data: & SVHN & \begin{tabular}[c]{@{}c@{}}Data\\ Free\end{tabular}  & CelebA & \begin{tabular}[c]{@{}c@{}}Tiny\\ imagenet\end{tabular}  & Imagenette & CIFAR-10      \\
\midrule
ZSDB3KD             & -             & 50.18              & -               & -                      & -                   & -                      \\
DFMS-HL (Ours)               & 84.83         & \textbf{84.51}              & 85.82           & 92.26                  & 90.06             & 81.98       \\
\bottomrule
\end{tabular}}
\vspace{-0.5cm}
\end{table}

\textbf{Comparison with Model Stealing methods.}
We compare our approach with the state-of-the-art data-free Model Stealing approaches~\cite{kariyappa2021maze, truong2021data} on CIFAR-10 in Table \ref{table:resnet_main}. We obtain an accuracy of 84.51\% by merely using synthetic samples in a completely data-free hard-label setting. We use a lower query budget of 8M, as compared to that of DFME and MAZE that require 20M queries for CIFAR-10. We further extend our attack to the soft-label black-box scenario (denoted as DFMS-SL in Table \ref{table:resnet_main}) where the softmax predictions of the victim model are available. In order to utilize the soft labels in a KD setting, we use the L1 loss formulation of DFME~\cite{truong2021data}, which computes the L1 distance between the victim and clone model's logits. Victim model logits are estimated by first taking log of the softmax output, followed by a mean correction. We use the same query budget of 20M and get a boost of almost 3\% using synthetic data and proxy data of 10 CIFAR-100 classes.

\textbf{Results on CIFAR-100.}
We perform experiments on CIFAR-100 (Table \ref{table:cifar100_main}) with CIFAR-10 \cite{addepalli2020degan,barbalau2020black} and synthetic data as proxy datasets using a ResNet-18 victim model of accuracy 78.52\%. DeGAN attains an accuracy of 75.62\% in a soft-label setting with access to the teacher gradients. DFMS-HL reaches a comparably close accuracy of 72.83\% using CIFAR-10 as the proxy and 43.56\% using synthetic data without any access to the victim model's gradients and only using hard labels. This shows that gradients from the clone model effectively act as a proxy to the victim model's gradients, for training the generator to generate diverse samples across all classes. 

\section{Ablation Study}
\label{sec:abla_study}
\begin{figure}
\centering
        \includegraphics[width=0.75\linewidth]{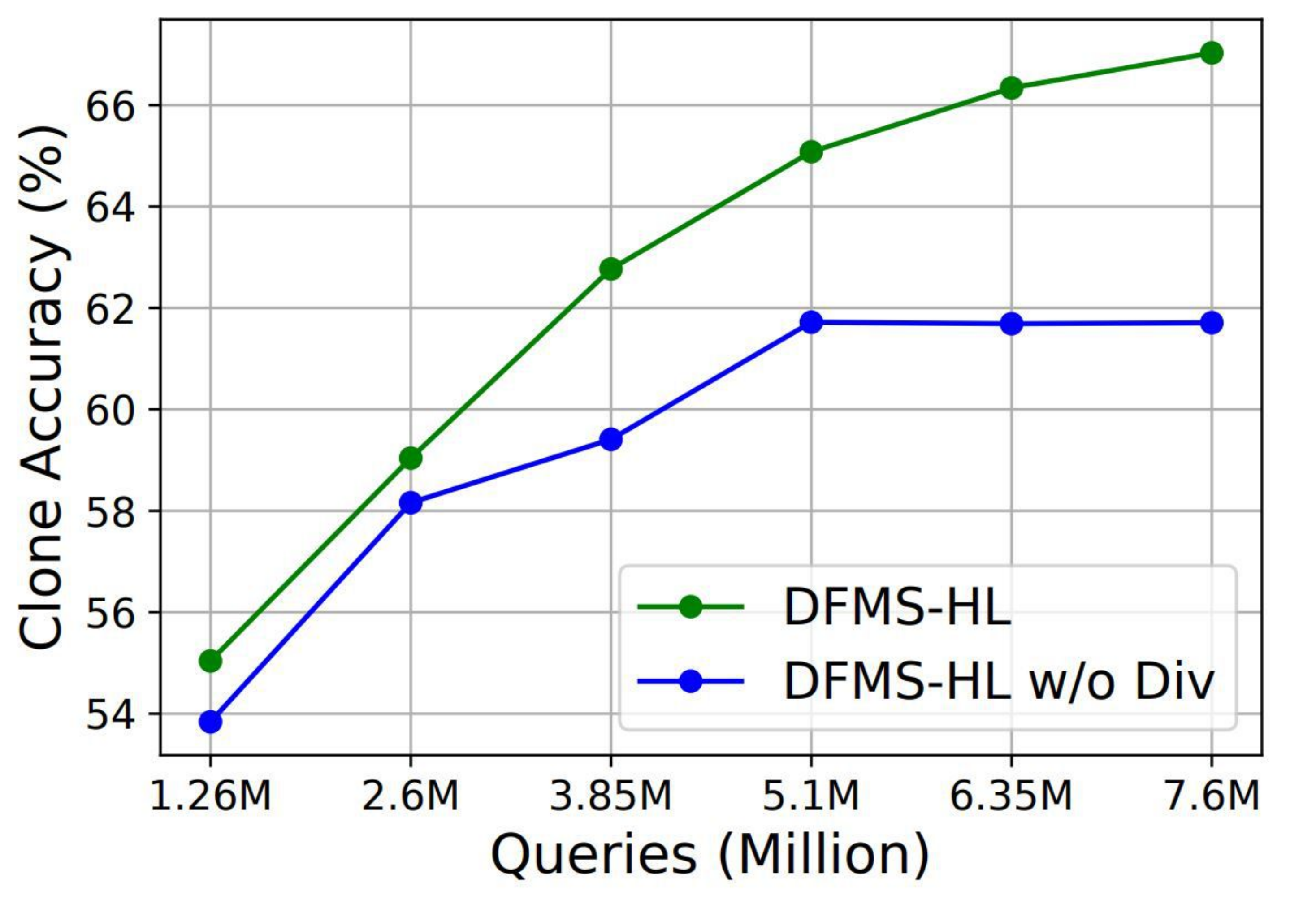}
        \caption{\textbf{Query Ablation on CIFAR-10 using synthetic images as proxy data:} Plot of clone model accuracy (\%) w.r.t. the number of queries. We achieve a significant boost of 6\% by using the class-diversity loss.}
        \label{fig:plot_queries}
        \vspace{-1em}
\end{figure}

\textbf{Query Budget:}
We analyze the impact of query budget on the accuracy of the clone model. Our approach achieves a good accuracy within a query budget of 7.6 million using synthetic data as proxy, with AlexNet as the victim model and AlexNet-half as the clone model on CIFAR-10. From Fig. \ref{fig:plot_queries} we observe that even with a small query budget of 1.26M, our method performs well and the accuracy saturates within 8M. We report the saturating accuracies in Tables \ref{table:cifar10_main} and \ref{table:resnet_main}. We use a query budget of 10M for the CIFAR-100 experiments (Table \ref{table:cifar100_main}) and 8M for the CIFAR-10 experiments (Tables \ref{table:cifar10_main} and \ref{table:resnet_main}). The class-diversity loss has a huge impact on the clone model accuracy as we observe a significant boost of 6\% using this. 

\begin{figure}
\centering
        \includegraphics[width=0.75\linewidth]{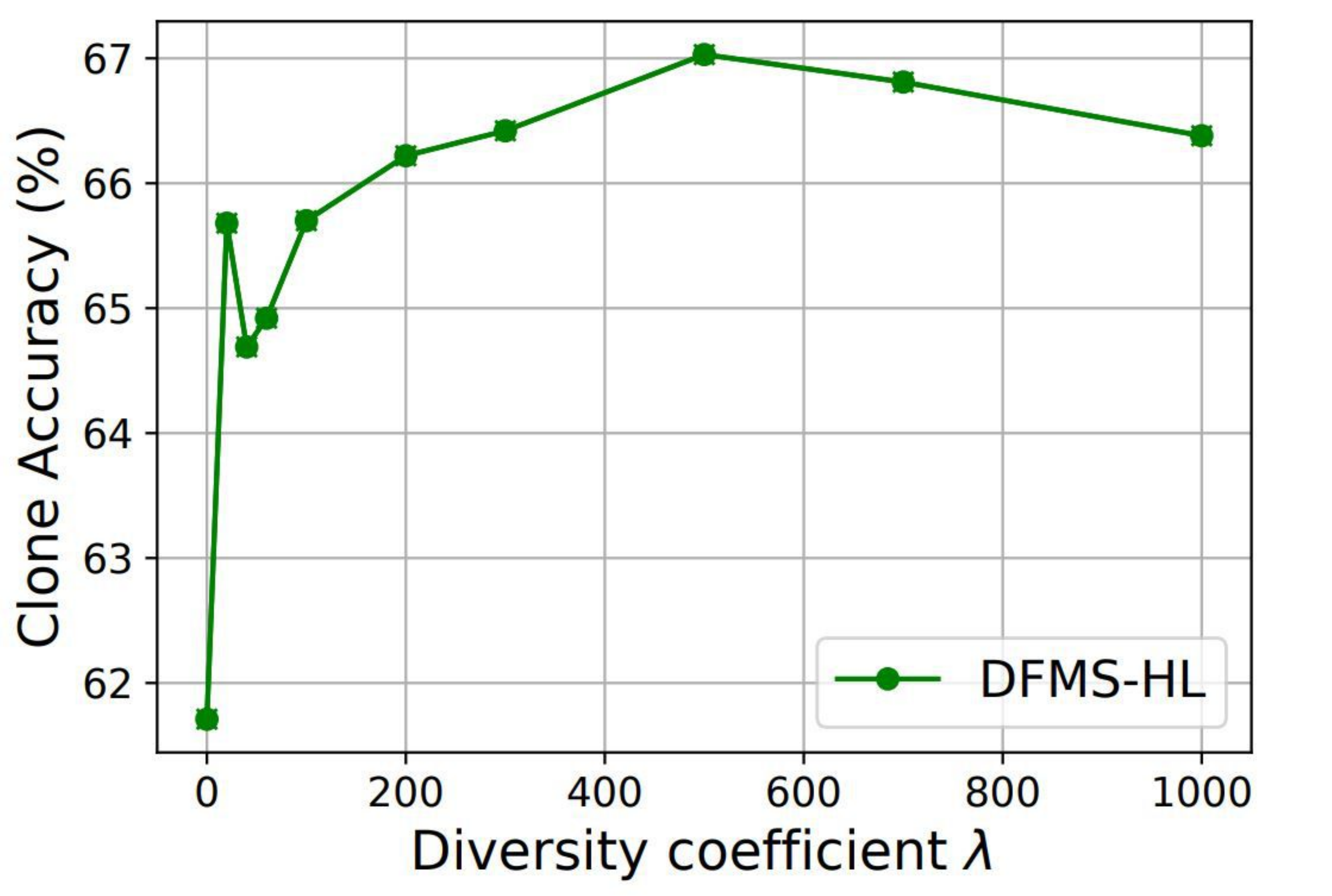}
        \caption{\textbf{Sensitivity Plot for Class-diversity Loss:}  Clone model accuracy is stable across a wide range of loss coefficients $\lambda_{div}$.}
        \label{fig:div_loss_syn}
        \vspace{-0.3cm}
\end{figure}

\textbf{Class Diversity Loss:}
We perform an ablation study by varying the coefficient of the diversity loss from 0 to 1000 in Fig. \ref{fig:div_loss_syn}.  We use synthetic data as proxy with CIFAR-10 as the original training dataset of the Victim model. We run the ablations for 150 epochs of training, which limits the queries to 7.6M. We find that the clone model accuracy is stable across a wide range of loss coefficients. We set $\lambda_{div}$ to 500 for CIFAR-10 and 100 for CIFAR-100.

\begin{figure}
\centering
        \includegraphics[width=0.72\linewidth]{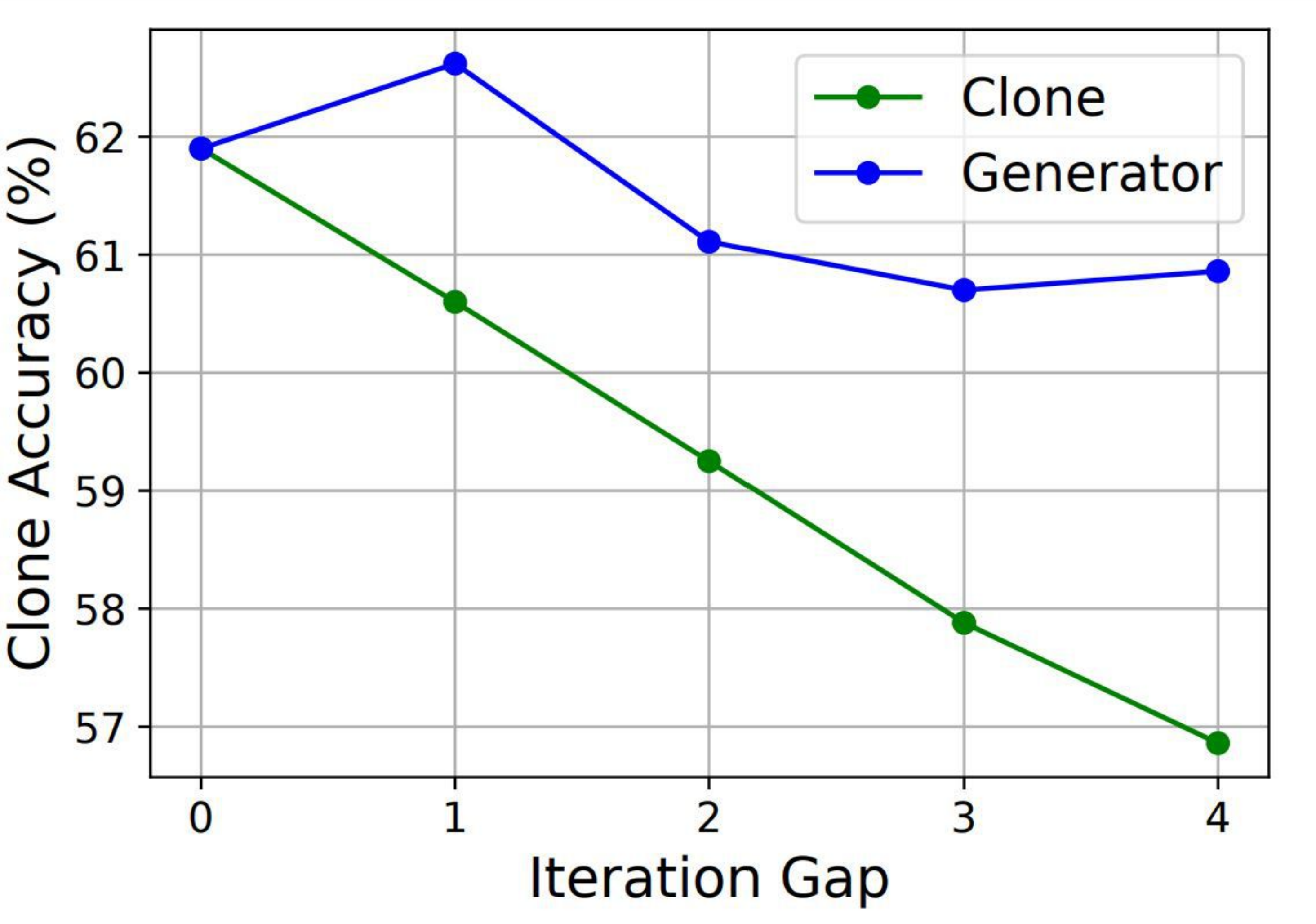}
        \caption{\textbf{Iteration Gap ablation:} Clone model accuracy plotted against iteration gap for training the clone and generator.}
        \label{fig:iter_gap}
        \vspace{-0.5cm}
\end{figure}

\textbf{Alternate training of Clone model and generator:}
The generator and clone model are trained once in every iteration. We check the impact of training each model after every $t$ iterations in Fig. \ref{fig:iter_gap}. We use synthetic dataset as proxy data and CIFAR-10 as the Victim training dataset, with 85 epochs of training for this ablation. We vary the iteration gap of training each model from 0 to 4. A gap of 0 indicates that the respective model is trained every iteration. The results show that increasing the iteration gap impacts the clone accuracy. We obtain a marginally better accuracy when the generator is trained in alternate iterations. We report our final results with iteration gap set to $0$ for both clone model and generator.

\begin{figure}
\centering
        \includegraphics[width=0.75\linewidth]{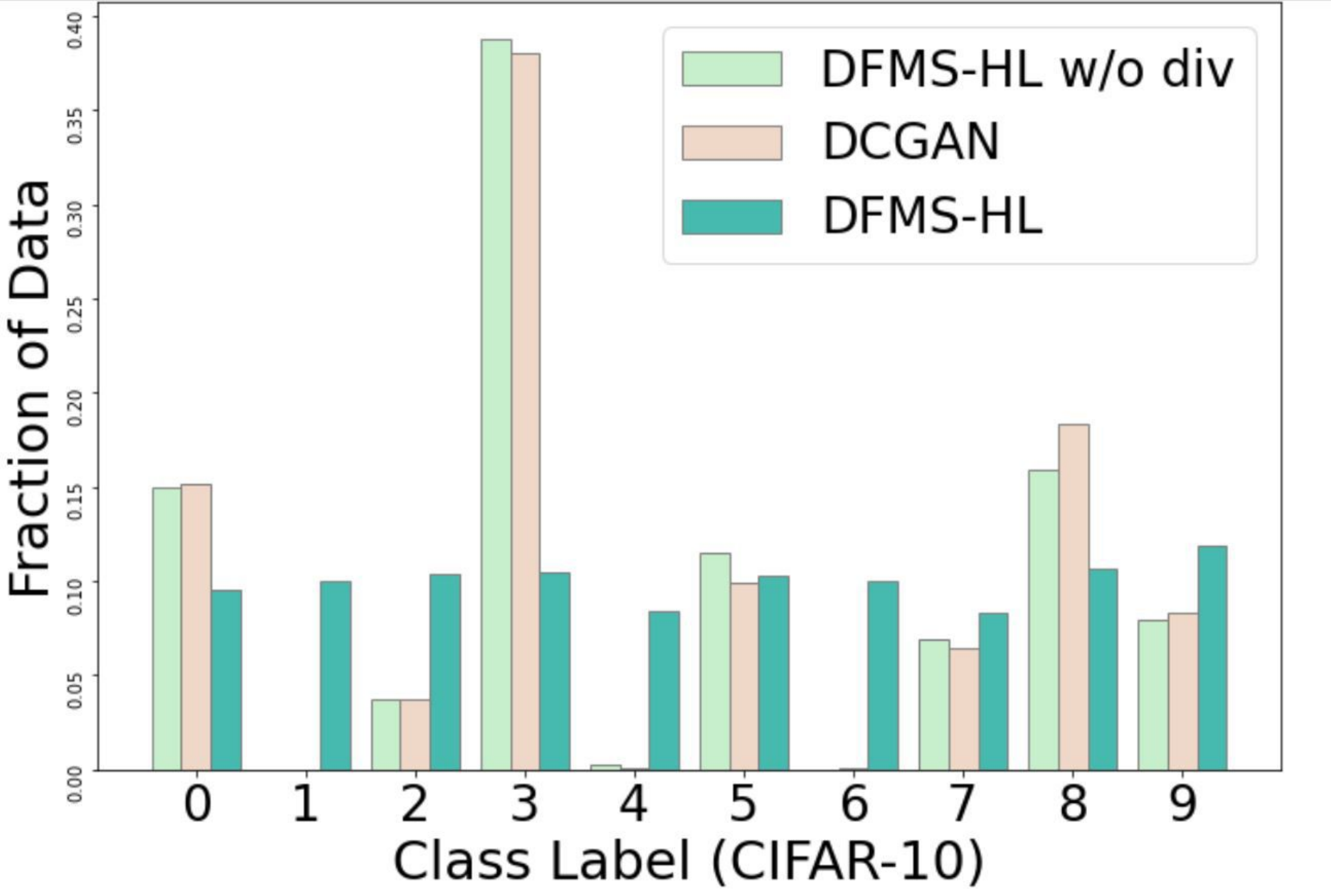}
        \caption{\textbf{Distribution of images across classes:} The images generated by DFMS-HL are distributed evenly across all classes. }
        \label{fig:class_dist}
        \vspace{-0.5cm}
\end{figure}

\textbf{Generation of Diverse Images:}
The DFMS-HL generator is initialized with a DCGAN generator at the start of the training process. As the training progresses, the generator learns to generate images distributed evenly across different classes of the victim model as shown in Fig. \ref{fig:class_dist}. We use synthetic images as the proxy data and CIFAR-10 as the Victim model training dataset, with AlexNet/ AlexNet-half as victim/ clone model architectures. The initial distribution of images generated using DCGAN is skewed, with very few samples in classes 1, 4 and 6. The distribution of images without the diversity loss is also skewed. Based on the plots, we note that the class-diversity loss has a huge impact in making the class distribution uniform.

\section{Conclusions}
In this paper, we propose an effective model stealing attack in a practical setting of having access to only hard-labels of a black-box victim model. Extensive experiments show that our method DFMS-HL performs better than the state-of-the art model stealing method ZSDB3KD at a $500\times$ lower query budget. 
We further show that our attack is effective in a completely data-free setting as well, that uses synthetically generated images to impose an image prior. We demonstrate the scalability of the proposed model stealing attack to CIFAR-100 within a low query budget, which has not been attempted in prior works. Our ablations reveal that the class-diversity loss plays a major role in achieving diversity in the generated images, boosting the clone model accuracy evenly across all classes.


Although our work describes methods of attacking the privacy of models through model stealing, the goal is indeed to create better awareness and understanding of the vulnerabilities of Machine Leaning models. This would in turn promote research towards the development of novel defenses against such attacks, leading to  a more robust ecosystem with increased security and privacy.

\section{Acknowledgements}
This work was supported by a project grant from MeitY (No.4(16) /2019-ITEA), Govt. of India and a grant from Uchhatar
Avishkar Yojana (UAY, IISC\_010), MHRD, Govt. of India. Sunandini Sanyal is supported by Prime
Minister’s Research Fellowship, and Sravanti Addepalli is supported by Google PhD Fellowship. We are thankful for the support.

{\small
\bibliographystyle{ieee_fullname}
\bibliography{egbib}
}

\appendix

\twocolumn[
  \begin{@twocolumnfalse}
\begin{center}
\textbf{\Large Supplementary Material:\\  
\qquad \\
Towards Data-Free Model Stealing in a Hard Label Setting}
\vspace{1cm}
\end{center}
 \end{@twocolumnfalse}
  ]

\section{Datasets}
We perform experiments using different proxy datasets similar to prior works~\cite{addepalli2020degan,barbalau2020black} to evaluate the effectiveness of our method DFMS-HL. This section contains a description of the different datasets that we used to evaluate our attack with CIFAR-10 as the true dataset.

\begin{itemize}
    \item \textbf{40-unrelated classes from CIFAR-100 \cite{addepalli2020degan}:} This consists of training data from CIFAR-100 belonging to non-overlapping classes with respect to CIFAR-10. The classes from the following categories are included: food containers, household electric devices, household furniture, large man-made outdoor things, large natural outdoor scenes, flowers, fruits and vegetables, trees.
    \item \textbf{10 random classes of CIFAR-100:} From the above 40 unrelated classes, we choose 10 classes randomly to demonstrate this setting. The classes used are : plate, rose, castle, keyboard, house, forest, road, television, bottle and wardrobe.
    
    \item \textbf{Synthetic Dataset:} We construct synthetic images which are far from the manifold of the training data distribution to simulate this setting. The images contain multiple overlapping shapes on top of a planar background. The creation of synthetic images is described in Sec. \ref{sec:synthetic_data}.
\end{itemize}

\subsection{Creation of Synthetic Dataset}
\label{sec:synthetic_data}
\begin{figure}
\centering
        \includegraphics[width=\linewidth]{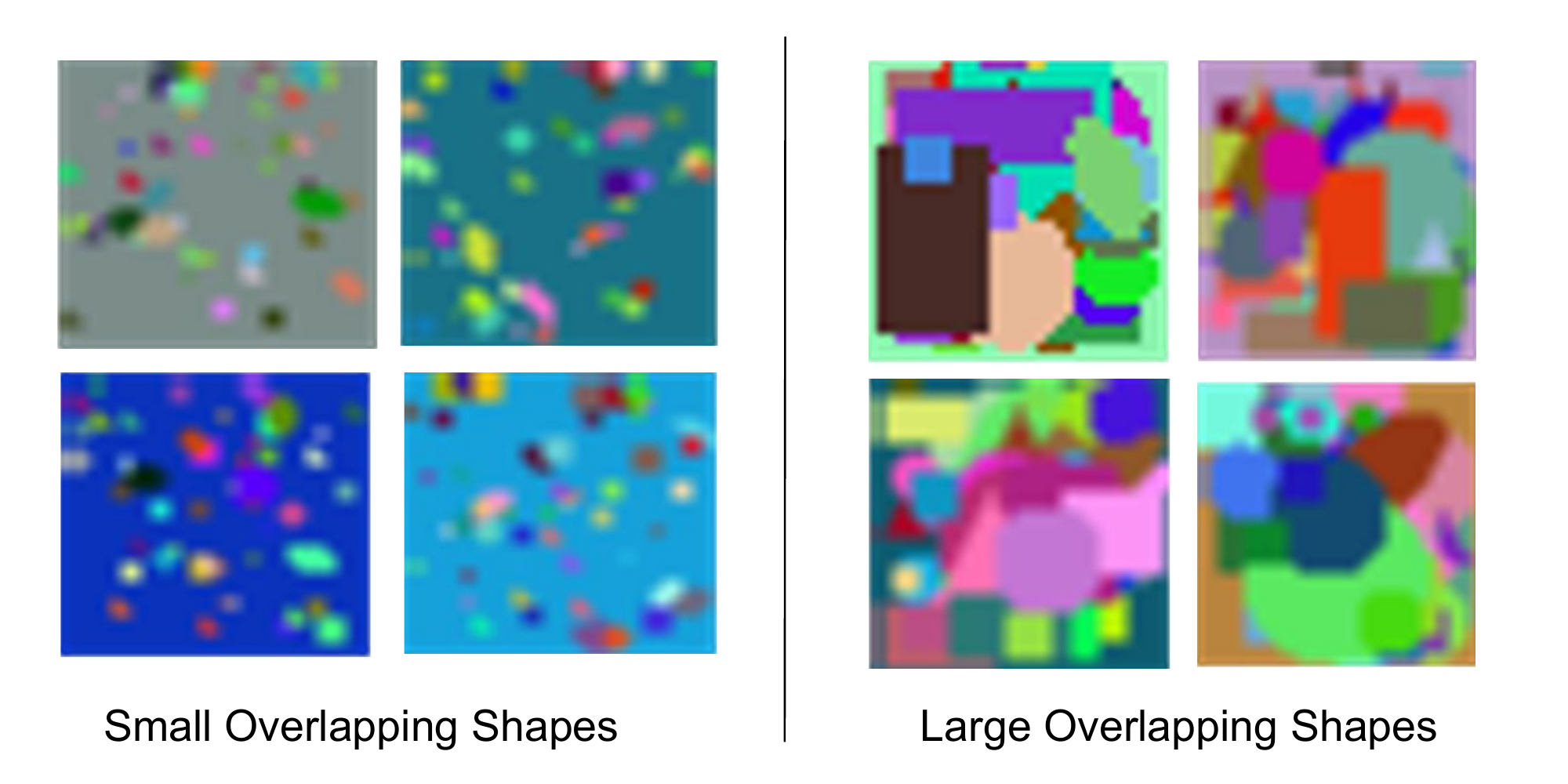}
        \caption{\textbf{Types of synthetic images used.} An equal share of large(right) and small(left) overlapping shapes on planar background used to train the clone model.}
        \label{fig:types_of_imgs}
\end{figure}

The algorithm to create a synthetic dataset is presented in Algorithm \ref{alg:synth_algo}. At first, randomly sampled shapes (triangle, rectangle, circle or ellipse) are generated at random locations in the image with a randomly sampled colour. The shapes are generated using python skimage module\footnote{\url{https://scikit-image.org/docs/stable/auto_examples/edges/plot_random_shapes.html}}. A total of 50K images are generated. We generate two kinds of images. The first variant contains large overlapping shapes with number of shapes in the image (num\_shapes) as 50 and the (min\_size, max\_size) of each shape as (20,50). The initial image generated is of size (100 x 100) which is scaled down to (32 x 32). The other variant contains textured images with (min\_size, max\_size) as (5,10) and num\_shapes=50 to get small overlapping shapes on top of a planar background. A random colour is sampled and assigned to the background pixels.  These images are then used to steal an ML model trained on CIFAR-10 and CIFAR-100. The generated images are shown in Fig. \ref{fig:types_of_imgs}. We share our dataset here\footnote{\url{https://drive.google.com/drive/folders/1CCMCYVRnvqZig9dYUYO_BupI8tImGZ2x}}.

\begin{algorithm}[t]
\caption{Algorithm for creating synthetic data}\label{alg:synth_algo}
\begin{algorithmic}
\Require Number of images to be generated $N_P$, num\_shapes, max\_size, min\_size
\While{$N_P \neq 0$}
    \State Generates shapes on an image of size (100 x 100), with parameters: num\_shapes, min\_size, max\_size
    \State Assign a random RGB colour to background pixels
    \State Perform blurring on the image using a 4 x 4 filter
    \State Resize image to (32 x 32)
    \State $N_P \gets N_P-1$
\EndWhile

\end{algorithmic}
\end{algorithm}

\section{Insights on Query Budget}
Chandrasekaran \etal \cite{chandrasekaran2020exploring} formulated the model extraction task as a query synthesis algorithm where an adversary $\mathcal{A}$ can ask for labels of the data $x$ which could be completely unrelated to the training data distribution. They show that, given a maximum query budget of $q_A(\epsilon,\delta)$ and a victim model $\mathcal{V}$ trained with a specific hypothesis $f^* \in \mathcal{F}$, there exists an adversary which implements an $\epsilon$-extraction attack with confidence $1-\delta$. Adversary $\mathcal{A}$ trains a clone model $\mathcal{C}$ with hypothesis $\hat{f}$ such that the following holds true.
\begin{equation}
    Pr[\mathcal{A}\ \text{trains} \ \hat{f} \text{and} \ \text{Err}(\hat{f}) \leq \epsilon] \geq 1-\delta
\end{equation}
where $Err(\hat{f}) = ||w^* - w||_2$, $w$ and $\hat{w}$ being the parameters of $\hat{f}$ and $f^*$, respectively. This shows that an adversary can implement a model stealing algorithm in a Query Synthesis scenario using active learning. Further, the authors \cite{chandrasekaran2020exploring} show that even when a victim employs
a randomized procedure for returning labels such that the upper bound on the probability of returning wrong labels $\rho_{D}(f^*)<\frac{1}{2}$, an adversary can implement an $\epsilon$-extraction attack with confidence $1-2\delta$ within the following query bound:
\begin{equation}
q =\frac{8}{(1 - 2 \rho_{D}(f^*))^{2}} q(\epsilon, \delta) ln \frac{q(\epsilon,\delta)}{\delta}
\end{equation}

\section{Experimental Details}
For the evaluation of DFMS-HL, we consider victim models trained on two datasets, CIFAR-10 and CIFAR-100. For each dataset, a victim model is trained upto a comparable accuracy of the teacher model used in prior works \cite{addepalli2020degan,barbalau2020black,wang2021zero}. The initial Clone model is trained with an SGD optimizer of momentum 0.9, learning rate of 0.1 and weight decay of $5 \times 10^{-4}$. We train the initial clone model for 200 epochs. The learning rate is reduced to 0.01 once the alternate training of clone and generator starts. After this, the clone is trained alternately till the query budget is exhausted. We use a cosine annealing scheduler to decay the learning rate across epochs. For the generator, a DCGAN architecture is trained with an Adam Optimizer and a learning rate of $2 \times 10^{4}$ with $(\beta_1, \beta_2)$ as (0.5, 0.999). We use NVIDIA GeForce GTX 1080 Ti and GeForce RTX 3090 to train our models. Our code takes a total training time of approximately 5 hours for CIFAR-10 and 10 hours for CIFAR-100 datasets on NVIDIA GeForce RTX 3090.

\section{Ablation Experiments}

\begin{figure}
\centering
        \includegraphics[width=\linewidth]{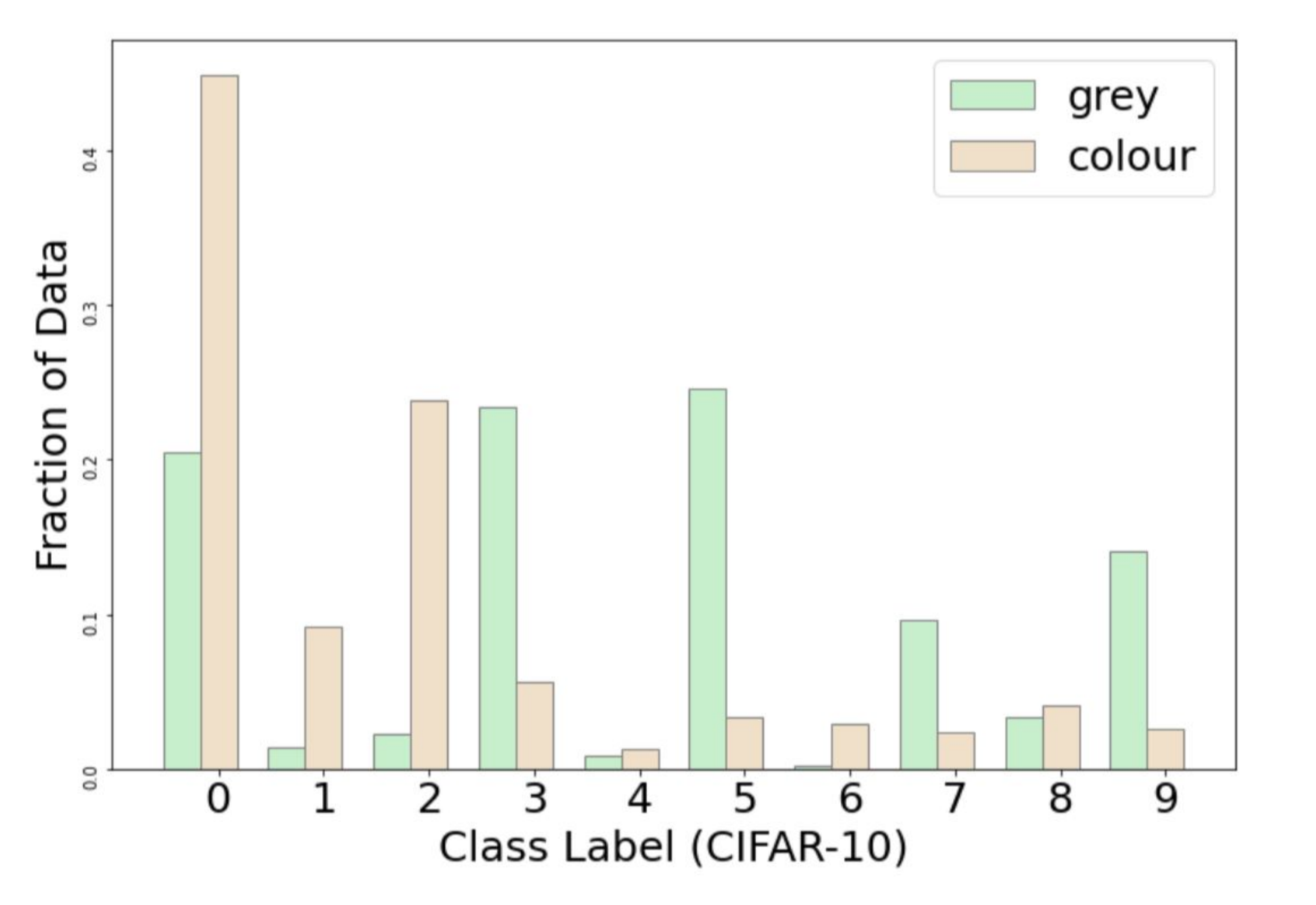}
        \caption{\textbf{Distribution of classes for grey vs colour images:} The grey synthetic images are more uniformly distributed across CIFAR-10 classes as compared to coloured images.}
        \label{fig:grey_vs_color}
\end{figure}

\subsection{Impact of Synthetic Data}
We tried two variants of the synthetic dataset. The first variant, ``Large overlapping shapes" contains multiple overlapping shapes on a planar background. The second variant ``Small overlapping shapes" contains multiple shapes of smaller size in an image. Each variant is shown in Fig \ref{fig:types_of_imgs}. We report results obtained by using each of these datasets individually and both combined in Table-\ref{table:synth_exp_1}. In this experiment, we use grey scale images for training. After combining the two datasets, we obtain a competent accuracy of 85.92\%. 

We use grey-scale and coloured images individually from the synthetic dataset and observe its impact on the clone model accuracy with an AlexNet victim network. We find that the grey images are well-distributed across multiple classes as shown in Fig. \ref{fig:grey_vs_color}. This makes grey images a better choice for initialization. In our method, we train a clone model with a mix of images from the proxy data and the generator to obtain a good initialisation. From our experiments, we observe that the initial clone model trained with grey-scale synthetic data achieves an accuracy of 44.57\%
and the one trained with coloured images has an accuracy of 37.31\%. This shows that grey-scale images lead to a better initialization for the clone model. Hence, we reported the final results of our method using grey-scale synthetic images. We also report the results of using the grey-scale and colour images individually for training in Table \ref{table:synth_exp_2} and observe that the final clone accuracy in both cases are comparable.

\begin{table}
\caption{\textbf{Impact of Synthetic Data}: Clone Model accuracy with different kinds of synthetic data images used, obtained on a ResNet-18 victim model of accuracy 93.65\%, with ResNet-18 as the clone architecture. }
\setlength\tabcolsep{3pt}
\centering
\resizebox{0.7\linewidth}{!}{
\label{table:synth_exp_1}
\begin{tabular}{lc}
\toprule
Type of Synthetic Data   & Clone Accuracy \\
\midrule
Large overlapping shapes & 80.34          \\
Small overlaping shapes  & 56.30           \\
Large + Small Combined   & \textbf{85.92} \\
\bottomrule
\end{tabular}}
\end{table}
\begin{table}
\caption{\textbf{Impact of Synthetic Data}: Comparison for grey vs coloured images used as proxy data, with AlexNet as the victim model of accuracy 80.18\% , trained on CIFAR-10, and AlexNet-half as the clone model.}
\setlength\tabcolsep{3pt}
\centering
\resizebox{0.7\linewidth}{!}{
\label{table:synth_exp_2}
\begin{tabular}{lc}
\toprule
Type of Synthetic Data        & Clone Accuracy \\
\midrule
Grey synthetic images         & 67.03          \\
Coloured synthetic images     & 65.84          \\
\bottomrule
\end{tabular}}
\end{table}

\subsection{Hyperparameter tuning}
The diversity loss plays a crucial role in ensuring that the distribution of images from the generator is class-balanced. The loss formulation of the generator with the class-diversity loss is shown below:
\begin{equation}
   \mathcal{L}_G  =  \mathcal{L}_{adv, fake} + \lambda_{div} \cdot  \mathcal{L}_{class\_div}  
\end{equation}
We show the impact of varying the class-diversity loss coefficient $\lambda_{div}$ in Table \ref{table:div_loss}. The true dataset is CIFAR-10 and the proxy dataset is 10 random classes from CIFAR-100. We use AlexNet as the victim architecture and train an AlexNet-half as the clone model for 500 epochs. We observe that as we increase the diversity loss coefficient, the clone model accuracy increases and reaches the maximum accuracy of 69.66\% at $\lambda_{div}$=500. We note that the proposed method is not sensitive to minor variations in the hyperparameter $\lambda_{div}$.

\begin{table}
\caption{\textbf{Impact of class-diversity loss coefficient $\lambda_{div}$}: Performance (\%) of the clone model on CIFAR-10 dataset trained using 10 random classes of CIFAR-100 as proxy, across variation in $\lambda_{div}$. The architecture of victim model is Alexnet and architecture of clone model is AlexNet-half.  The proposed method is not sensitive to minor variations in $\lambda_{div}$.}
\setlength\tabcolsep{3pt}
\centering
\resizebox{0.7\linewidth}{!}{
\label{table:div_loss}
\begin{tabular}{cc}
\toprule
Diversity Loss Coefficient & Clone Accuracy  \\
\midrule
100                        & 69.29          \\
200                        & 69.59          \\
300                        & 69.42          \\
\textbf{500}               & \textbf{69.66} \\
700                        & 69.54          \\
1000                       & 69.13         \\
\bottomrule
\end{tabular}}
\end{table}

\subsection{Impact of Clone architecture}
In a practical scenario of Model Stealing, the architecture of the victim model is unknown to the attacker. Hence, we aim to stage a successful attack in a completely black-box condition. To evaluate the effectiveness of the attack in different scenarios, we perform an ablation experiment to see if the choice of the clone model architecture impacts the success of the attack. The clone model achieves a high accuracy of 83.37\% using 10 random classes of CIFAR-100 when the same ResNet-18 architecture is used for both the victim and the clone. However, using a deeper CNN model such as GoogleNet gives a boost to the clone accuracy as shown in Table \ref{table:surrogate_exp}. We get lower clone accuracy for shallower networks such as AlexNet-half and VGG-11. Hence, we observe that it is beneficial for an adversary to use a deeper CNN architecture for capturing complex features from the victim model using proxy data.

\begin{table}
\caption{\textbf{ Impact of clone architecture on clone accuracy}: Clone Accuracy improves with a deeper CNN network}
\setlength\tabcolsep{3pt}
\centering
\resizebox{0.7\linewidth}{!}{
\label{table:surrogate_exp}
\begin{tabular}{lc}
\toprule
Clone Model Architecture & \multicolumn{1}{l}{Clone Accuracy} \\
\midrule
ResNet-18                & \textbf{83.37}                     \\
AlexNet                  & 79.37                              \\
AlexNet\_half            & 62.64                              \\
VGG-11                   & 74.59                              \\
VGG-19                   & 78.85                              \\
GoogleNet                & 84.50 \\                             
\bottomrule
\end{tabular}}
\vspace{-0.1cm}
\end{table}

\subsection{Impact of Discriminator}
\begin{figure*}
\centering
        \includegraphics[width=\linewidth]{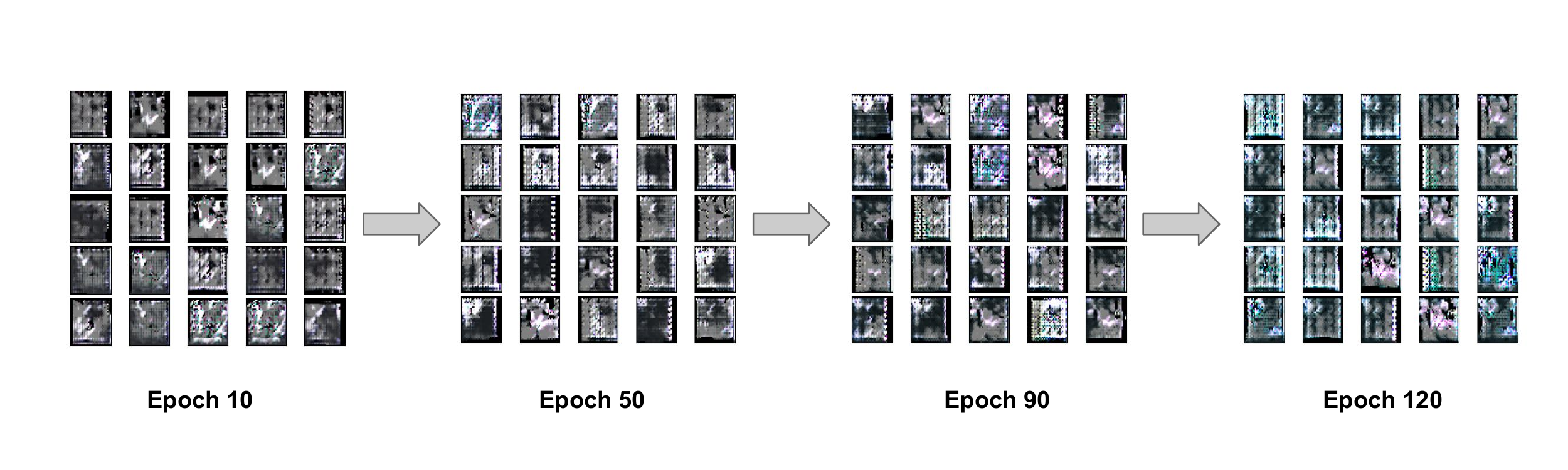}
        \caption{\textbf{Output of DFMS-HL after disabling the discriminator.} The images converge to degenerate cases after few epochs of training. Synthetic data is used as proxy data with an AlexNet victim model trained on CIFAR-10 and clone model as AlexNet-half.}
        \label{fig:w/o_dis}
\end{figure*}
The discriminator is an essential component of our approach. Across training epochs, the discriminator learns to differentiate between proxy data and fake images produced by the generator. We conduct an ablation experiment by disabling the discriminator updates. We use CIFAR-10 as the true dataset and synthetic data as the proxy dataset for this experiment. For Alexnet as victim model and AlexNet-Half as clone model, DFMS-HL  attains an accuracy of 67.03\%. After disabling the discriminator, the clone accuracy drops to 57.06\% and the images look degenerate as shown in Fig. \ref{fig:w/o_dis}. Hence, the discriminator also plays a crucial role in maintaining the distribution of images. 

\begin{table}
\caption{\textbf{Impact of L1 loss formulation on DFMS-SL (Soft-Label Setting)}: Clone Model accuracy increases by 3\% after using L1-loss as compared to standard KL-divergence loss. Synthetic data is used as proxy for a ResNet-34 victim model trained on CIFAR-10 and  ResNet-18 used as Clone model.}
\setlength\tabcolsep{3pt}
\centering
\resizebox{0.7\linewidth}{!}{
\label{table:soft_label}
\begin{tabular}{ccc}
\toprule
\textbf{Method}      & \textbf{Teacher Acc} & Synthetic                \\
\midrule
DFME                 & 95.5                 & 88.10                    \\
DFMS-SL(L1 loss)     & 95.5                 & 91.24                    \\
DFMS-SL(KL-div loss) & 95.5                 & 88.40\\
\bottomrule
\end{tabular}}
\end{table}

\begin{table}
\caption{\textbf{SVHN as Proxy Data ablation}: DFMS-HL achieves an accuracy of 84.83\% using SVHN as Proxy data for a ResNet-34 victim model trained on CIFAR-10. ResNet-18 used as Clone architecture.}
\setlength\tabcolsep{3pt}
\centering
\resizebox{1.0\linewidth}{!}{
\label{table:svhn}
\begin{tabular}{ccccc}
\toprule
\textbf{Method} & Synthetic & CIFAR-100 (40C) & CIFAR-100 (10C) & SVHN  \\
\midrule
DFME                           & 88.10     & 88.10           & 88.10           & 88.10 \\
DFMS-HL (Ours)                 & 84.51     & 92.06           & 85.53           & 84.83 \\
\bottomrule
\end{tabular}}
\end{table}

\begin{figure*}
\centering
        \includegraphics[width=\linewidth]{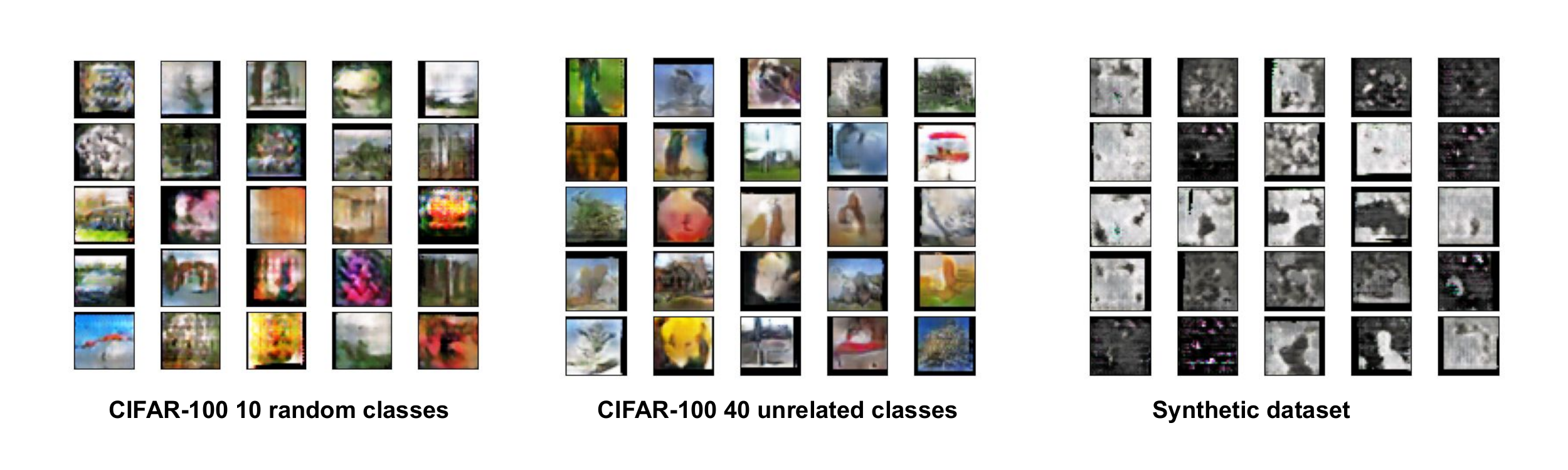}
        \caption{\textbf{DFMS-HL generator images.} The images generated by DFMS-HL generator for CIFAR-100 10 random classes, 40 unrelated classes and synthetic data as proxy for an AlexNet victim model of accuracy 80.18\% trained on CIFAR-10 and clone model as AlexNet-half.}
        \label{fig:gan_imgs_1}
\end{figure*}

\begin{figure*}
\centering
        \includegraphics[width=\linewidth]{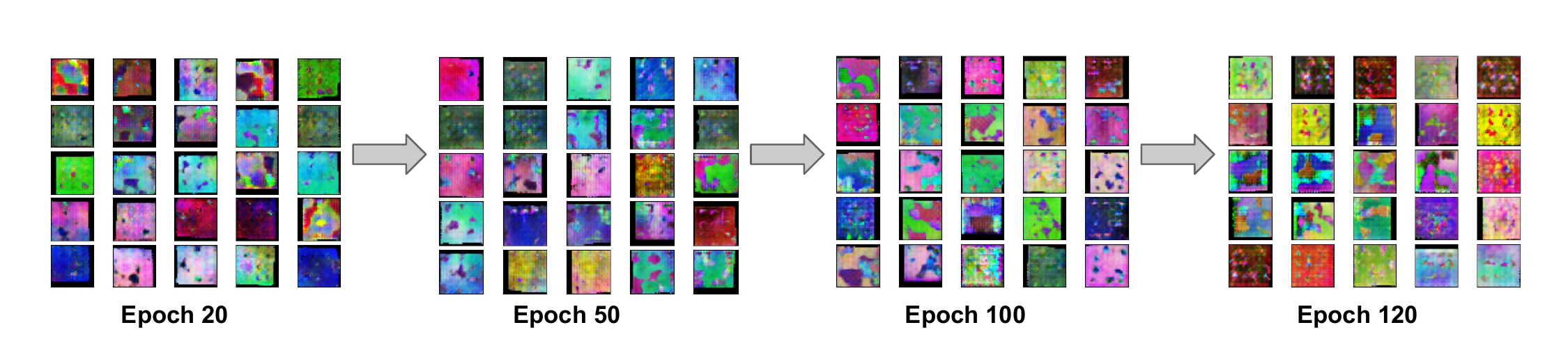}
        \caption{\textbf{DFMS-HL generator images.} The images generated by DFMS-HL generator using synthetic colour dataset as proxy for an AlexNet victim model of accuracy 80.18\% 
        trained on CIFAR-10 and clone model as AlexNet-half.}
        \label{fig:gan2_imgs}
\end{figure*}

\begin{figure*}
\centering
        \includegraphics[width=\linewidth]{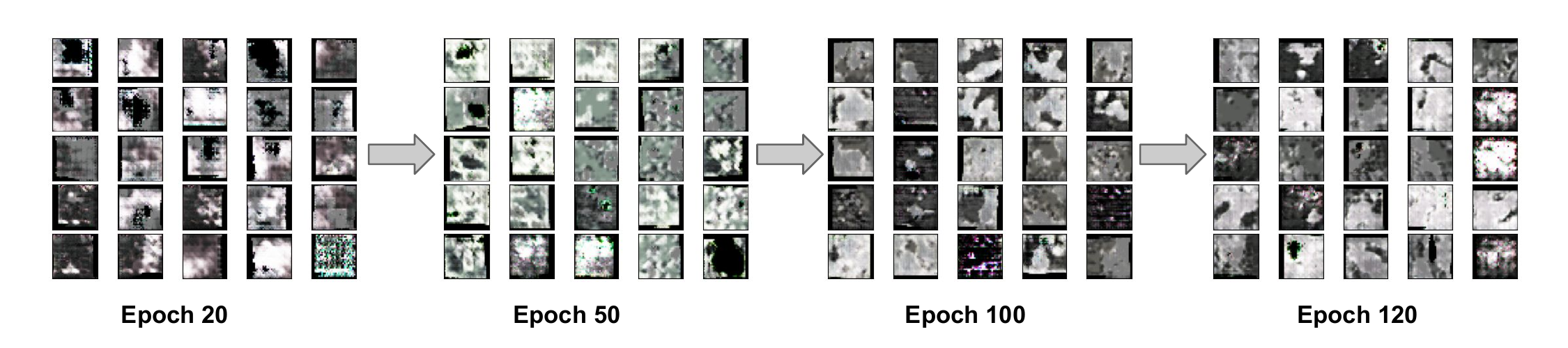}
        \caption{\textbf{DFMS-HL generator images.} The images generated by DFMS-HL generator using grey-scale synthetic images as proxy for an AlexNet victim model of accuracy 80.18\% trained on CIFAR-10 and clone model as AlexNet-half.}
        \label{fig:gan_imgs_3}
\end{figure*}

\subsection{Impact of L1 loss in DFMS-SL}
Prior works on Knowledge Distillation~\cite{,hinton2015distilling,lopes2017data,nayak2019zero} train a student model using a KL-divergence loss between the student and teacher predictions. Let $\mathcal{V}_i(x)$ and $\mathcal{C}_i(x)$ be the output of class $i$ (out of $K$ classes) of the victim and clone models respectively. The KL divergence loss is written as follows,

\begin{equation}
    \displaystyle \mathcal{L}_{KL} = \sum_{i=0}^{K} \mathcal{V}_i(x) log \left[ \frac{\mathcal{V}_i(x)}{\mathcal{C}_i(x)} \right]
\end{equation}

The DFME approach~\cite{truong2021data} used an L1 loss formulation where they consider the L1 difference between the logits of the clone and the victim model. The logits are estimated by first taking log, then subtracting the mean of the predictions from it. The loss formulation is written as follows,

\begin{equation}
    \displaystyle \mathcal{L}_{l1} = \sum_{i=0}^{K} |~ \mathcal{V}_{i}^{ logits}(x) - \mathcal{C}_{i}^{ logits}(x) ~|
\end{equation}
where,
\begin{equation}
    \mathcal{V}_{i}^{logits}(x) = log \mathcal{V}_{i}(x) - \frac{1}{K} \sum_{j=1}^{K} log \mathcal{V}_{j}(x)
\end{equation}

We evaluate our approach in the soft-label setting with the two loss functions of L1 loss and KL-divergence loss as shown in Table \ref{table:soft_label}. We observe an improvement in the clone accuracy using synthetic data by 3\% by using L1 loss for distillation.

\subsection{Using unrelated data as the Proxy Dataset}
The amount of relatedness between the proxy data and true data is an important factor that influences the success of model stealing. We perform an ablation study using SVHN as the proxy dataset to steal a model originally trained on CIFAR-10. Since SVHN is a completely unrelated to CIFAR-10, it is indeed a difficult setting. Our method DFMS-HL attains a clone accuracy of 84.83\% in this setting. This shows our attack is strong enough to work across a wide range of unrelated proxy datasets.

\section{GAN generated Images}
The images generated from the DFMS-HL GAN are shown in Fig. \ref{fig:gan_imgs_1}, \ref{fig:gan2_imgs} and \ref{fig:gan_imgs_3}. Initially, the generator starts generating images which closely resemble the proxy data. In the synthetic data experiments (Fig.\ref{fig:gan2_imgs} and \ref{fig:gan_imgs_3}), as the training progresses, we observe that the shapes start merging with each other and start looking more continuous in nature. This makes the image look close to real images which have an object in front of a background. This shows that the generator starts capturing properties of the true training data distribution, as they look more intuitive than the original synthetic images. This helps the clone model learn intrinsic properties of the victim's training data.

\section{Limitations and Future Directions}
One of the crucial factors of a successful model stealing attack is its query budget. Our approach has reduced the number of queries required to 8 million, which is $\sim500\times$ lesser than the query budget used by past methods of model stealing and knowledge distillation. We believe that reducing the query budget further would be an interesting area for future research. Another limiting factor for an adversary is the lack of relevant training data. Our approach addresses this limitation to quite an extent, as we showcase promising results in a limited data scenario by just using synthetic images. We believe that our approach would pave the way to address these limitations and develop stronger attacks and defenses in the area of hard-label model stealing.

\end{document}